\documentclass[runningheads]{llncs}
\usepackage{graphicx}
\usepackage{hyperref}
\usepackage[utf8]{inputenc}
\usepackage{times}
\usepackage{xcolor}
\usepackage{soul}
\usepackage{lineno,hyperref}
\usepackage{booktabs} 
\newcommand{\code}[1]{\textsf{\small #1}}

\begin{document}
\title{Causal Inference for Social Discrimination Reasoning}
%
%
\author{Bilal Qureshi\inst{1} \and
Faisal Kamiran\inst{2} \and
Asim Karim\inst{3} \and
Salvatore Ruggieri\inst{4} \and
Dino Pedreschi\inst{4}
}
\authorrunning{B. Qureshi et al.}
%
\institute{Addo.ai, Pakistan \email{bilal@addo.ai} \and
Information Technology University of the Punjab, Pakistan
\email{faisal.kamiran@itu.edu.pk} \and
Lahore University of Management Sciences (LUMS), Pakistan\\
\email{akarim@lums.edu.pk} \and
Universit\`a di Pisa and ISTI-CNR, Italy
\email{\{ruggieri,pedre\}@di.unipi.it}}
\maketitle              
\begin{abstract}
The discovery of discriminatory bias in human or automated decision making is a task of increasing importance and difficulty, exacerbated by the pervasive use of machine learning and data mining. Currently, discrimination discovery largely relies upon correlation analysis of decisions records, disregarding the impact of confounding biases. We present a method for \emph{causal} discrimination discovery based on \emph{propensity score analysis}, a statistical tool for filtering out the effect of confounding variables. We introduce causal measures of discrimination which quantify the effect of group membership on the decisions, and highlight causal discrimination/favoritism patterns by learning regression trees over the novel measures. We validate our approach on two real world datasets. Our proposed framework for causal discrimination has the potential to enhance the transparency of  machine learning  with tools for detecting discriminatory bias both in the training data and in the learning algorithms.

\end{abstract}

\section{Introduction}

\label{intro}
Discrimination can be instilled in automated decision-making systems at several places: it can be learned from training data; it can be a side-effect of biased algorithms; or, it can arise from ``dumb-smart" programs that are used in contexts they are not designed or tested for. 
Understanding and mitigating forms of bias in data/web mining systems, and  discrimination against protected social groups in particular, has been a growing research field in the last years, due to mounting awareness and concerns \footnote{See e.g.,~the 
AI Now report on \textit{Discriminating Systems}, April 2019 (\href{https://ainowinstitute.org/discriminatingsystems.pdf}{https://ainowinstitute.org}).}.

The large body of research on discrimination and fairness ~\cite{Barocas2016,RR2014}, however, largely ignores that reasoning on observational data through correlation analysis may produce biased results, due to {\em confounding variables}. 
Consider, e.g., the well-known UC Berkeley gender bias case~\cite{Bickel1975}: the admission outcomes for the fall of 1973 seemed to show that men were more likely than women to be admitted, a large difference indicating a potential discrimination. But dis-aggregating the numbers by the different departments, it emerged a ``small but statistically significant bias in favor of women'', due to fact that women tended to apply to competitive departments, whereas men tended to apply to less-competitive departments. Hence, the department confounds or biases the observed difference in admittance rates for males and females. The department here is a confounding variable because it is correlated both with the admittance decision and the gender of an applicant. It can also be thought of as an explanatory attribute that explains the overall preference for males, revealing that, in this case, gender discrimination did not occur. In a given analysis, there may be several confounding variables, and it is essential that their effect is minimized while estimating discrimination. Once accounted for confounding variables, we can reason on the \emph{causal} effect of, say, gender, on the admittance decision.

Discovering and understanding causal influences among variables is a fundamental goal of any data analysis process. In general, randomized control trials (RCT) are the gold-standard for inferring such influences. This is the case, for example, of online controlled experiments~\cite{DBLP:reference/ml/KohaviL17}, 
which are widely used to test for effectiveness of website personalization and of advertising strategies. 
Unfortunately, in discrimination discovery processes, RCTs 
are rarely possible or practical. In the vast majority of cases only observed data are available, along with domain expert's knowledge. Nonetheless, causal influences are more likely to be acceptable in court of law than mere correlations or statistical test of hypothesis alone~\cite{Foster2004}. This is why recent works, discussed in Section 5, propose to reason on causal discrimination based on a network of causal relations among variables, specified with the intervention of a domain expert. However, this approach is unpractical in real high-dimensional problems, such as those arising in web analytics, as the difficulty of estimating causal inter-dependencies scales super-linearly with the number of features, quickly becoming intractable.

In this paper, we introduce an approach for causal discrimination discovery and understanding that does not require extensive expert's intervention, based on {\em propensity score analysis} (PSA). The propensity score is the probability of membership to the protected group (e.g., females) given the individual's or group's attributes. PSA provides a principled way to `filter out' the explainable effect of confounding variables and to quantify the causal effect of the protected attribute (e.g., gender). In the UC Berkeley example above, e.g., PSA accounts for the effect of the departments to which candidates apply and confirms that there is no evidence of women discrimination. We adopt propensity score weighting to balance the distributions of protected and unprotected group instances. Based on the modified distribution, we quantify the degree of discrimination or favoritism of each instance. More generally than a standard application of PSA, we learn a regression tree to understand the characteristics of instances falling under different discrimination/favoritism bands. We evaluate our approach on real-world datasets. The results confirm that the causal effect of group membership can be higher or lower than the effect computed by ignoring the influence of confounding variables. Thus, our approach promises defendable discrimination analyses in practice. 

Our method has relevant impact towards data-driven machine learning processes aimed at developing decision support systems. \textit{First}, causal discrimination discovery provides a solid reasoning framework to detect biases contained in the training data, which may be inherited by the learned decision model. In turn, this prevent constructing black-box systems that incorporate discriminatory rules (see, e.g.,~\cite{Dresseleaao5580}) and whose decisions, by definition of black-boxes, cannot be easily interpretable (see~\cite{guidotti2018survey}). \textit{Second}, our proposed regression tree model for estimating the discrimination risk score offers a method for understanding the causal structure of the biases, and therefore repairing or sanitizing the training data for avoiding unwanted discrimination in the learned model. \textit{Third}, similar considerations apply to the validation of a learned model on test data, in that causal discrimination discovery may be applied to the test examples labeled by the learned model, to assess any potential algorithmic bias introduced by the adopted machine learning technique. Thus, our proposed causal discrimination framework may help improving the quality of the data/web mining process, enhancing trust and transparency in the outcomes.

This paper is organized as follows. First, the related work is discussed in Section~\ref{sc:rel_work}.
Section~\ref{causalindividualdiscrimination} introduces basic tools for discrimination analysis and presents the propensity score approach for causal discrimination against individuals, which is extended to causal discrimination against groups in Section~\ref{causaldiscriminationdiscovery}. Experiments are reported in Section~\ref{sc:exp}. Finally, Section~\ref{conclusions} summarizes our contribution and future work.

\section{Related Work}
\label{sc:rel_work} 

Early approaches for discrimination discovery have primarily relied upon simple statistical measures of correlation between the protected group attribute and the decision attribute~\cite{DBLP:journals/datamine/Zliobaite17}. Albeit causation has been recognized as significant in the legal circles~\cite{Foster2004} and in medical research~\cite{Grimes2002} for a while, the perils of correlation analysis have been pointed out in the data mining area only recently. 
When  decisions are made automatically, e..g, by a machine learning model, the problem of preventing discriminatory decisions is known as \textit{fairness}. See~\cite{Berk2017,DBLP:conf/icse/VermaR18} for surveys of quantitative measures of fairness/discrimination. 
A recent overview on causal approaches to fairness is~\cite{DBLP:journals/corr/abs-1805-05859}.
Counterfactual fairness~\cite{DBLP:conf/nips/KusnerLRS17}, in particular, requires that the prediction of a machine learning model for an individual to be the same as that in the counterfactual world,
where the individual had belonged to a different social group.
More closely to our approach,
propensity score based \textit{stratification} was adopted in~\cite{Calders2013controlling} to filter out the effect of confounding variables before learning fair linear regression models. 
In our work, however, we employ propensity score \textit{weighting} and address the problem of discrimination discovery and quantification from data, rather than the fairness problem in the design of machine learning models. Thus, direct comparison can be made with causal approaches for discrimination discovery.

In~\cite{DBLP:journals/ijdsa/BonchiHMR17} the causal structure of attributes (a Bayesian Network) is learned from data which is then analyzed to identify patterns of discrimination or favoritism. The construction of the causal network is however impractical for a large
number of attribute-value pairs, and it also requires human intervention in providing a temporal order between pairs. 
\cite{DBLP:conf/ijcai/ZhangWW16}  extends, as we do, the situation testing approach of~\cite{DBLP:conf/kdd/ThanhRT11} to causal reasoning. They adopt a Causal Bayesian Network, which, again, must be specified by a domain expert, to adjust the distance measure between tuples to include only the directly causal attributes. Our approach, instead, relies on the re-weighting of tuples to uniform the distribution in the protected and unprotected groups, without requiring human intervention. 
On the other hand, graphical and structural 
causal models~\cite{DBLP:conf/aaai/ZhangB18,DBLP:journals/ijdsa/ZhangW17}, 
are able to distinguish and to explain the effects of direct, indirect, and spurious discrimination. I.e.,  the causal influence of the protected attribute (direct discrimination), of correlated attributes (indirect discrimination), and of common causes (spurious discrimination) on the decision attribute. In this paper, instead, we deal only with direct discrimination. Finally, some approaches (see e.g.,~\cite{DBLP:conf/kdd/ZhangWW17}) are extended to tackle training data sanitization. While this is out of the scope of this paper, the method of the original paper~\cite{DBLP:conf/kdd/ThanhRT11} can be extended to data sanitization w.r.t.~our causal risk difference measure. 

In addition to causal graphical models, several other methods are available from the statistical~\cite{Agresti2002} and econometric literature~\cite{FORTIN2011} to control for confounders in observational data, including restriction, matching, stratification,  decomposition, regression, ANCOVA, etc.
Propensity score analysis (PSA) was originally proposed in~\cite{Rosenbaum83} and it has been widely applied in economics, medicine, and social sciences. See~\cite{Austin2011,GuoFraser2015} for an overview.
Uses of propensity scores include: matching homogeneous samples 
(matches are pairs/sets of individuals 
from treatment and control groups with similar characteristics), stratification or sub-classification 
(strata are sets of individuals whose propensity score in a same bin interval), 
weighting cases 
to account for the
difference in the distribution of the observed covariates between treatment and control
groups. Let us relate these notions to our approach. \textit{First}, matching is a form of nearest-neighbor search, but done on the uni-dimensional space of propensity scores, and for tuple(s) of different groups. 
We also build a sub-set of available data in the neighborhood of a tuple for assigning a discrimination label to the tuple (individual discrimination), according to the legal practice of situation testing~\cite{Bendick07,DBLP:conf/kdd/ThanhRT11}.
Our neighborhood definition (called kset), however, works in the multi-dimensional space of tuple attributes and looks for tuples of any group. \textit{Second}, regarding weighting, our key definition of causal risk difference (Definition~\ref{def:rdc}), turns out to be an estimator of the average treatment effect for the treated (ATT)~\cite{MorganTodd2008}. In our context, the ATT is the expected difference in negative decision (also known as potential outcome) among individuals of the protected group (also known as treatment group). Hence causal risk difference is a counter-factual measure of suffered discrimination by the specific tuple under consideration. In classical propensity score weighting, instead, the ATT is considered with regard to the whole available data. \textit{Third}, standard PSA targets a single contingency table for the whole dataset at hand, so considering discrimination at aggregate, level. Our approach extends standard PSA to extract a  description of the groups with the highest discrimination by learning regression trees over the dataset enriched with an attribute measuring causal individual discrimination. Paths in the regression trees unveil specific conditions of  (discovered) discrimination.

Finally, we acknowledge that PSA, and a fortiori our approach, suffers from some limitations. \textit{First}, it cannot handle unknown or unmeasured covariates in the data, unless they are correlated with measured ones. Such a limitation, however, is also true for causal approaches~\cite{DBLP:conf/uai/KilbertusBKWS19,DBLP:conf/ijcai/Wu0W19}. \textit{Second}, it works on assumption of ``strong ignorability'', namely that 
protected attribute (treatment)
is independent of the decision (potential outcome) conditional on the observed covariates. This is not easy to establish~\cite[Section 11.3.5]{Pearl2009book}, and, if it does not hold, bias reduction is not guaranteed. \textit{Third}, the precise assumption that the protected attribute of race can be considered as ``treatment'' in the context of counterfactual causal analysis is being debated in the legal literature~\cite{Kohler2017}.

\begin{figure}[t]
\begin{center}\small
\begin{tabular}{c|cc|c}
\multicolumn{1}{c}{} &  \multicolumn{2}{c}{decision}\\
\cline{2-3}
 \multicolumn{1}{c}{group} & \multicolumn{1}{c}{$\ominus$} & \multicolumn{1}{c}{$\oplus$} \\
  \hline
  protected & $a$ & $b$ & $n_1$\\
  unprotected & $c$ & $d$ & $n_2$\\ \hline
  & $m_1$  & $m_2$   & $n$\\
\end{tabular}\\[2ex]
\noindent
\[ p_1 = a/n_1 \quad  p_2 = c/n_2  \quad \mathit{RD} = p_1 - p_2 \]
\end{center}
\vspace{-2ex} \caption{4-fold, contingency table for
$\mathit{kset}_{\mathcal R}({\bf r}, k)$.} \label{symbolic}
\end{figure}

\section{Causal Individual Discrimination} \label{causalindividualdiscrimination}

\textbf{Individual Discrimination.}
Consider an individual ${\bf r}$ from a protect\-ed-by-law group (from now on, protected group) suffering from a negative decision, i.e.,~$\mathit{dec}({\bf r}) = \ominus$, and the central question: was ${\bf r}$ discriminated? A \textit{prima-facie} answer is to contrast \textit{ex-post} the decisions made for individuals sharing similar features in a way inspired by the legal approach of \textit{situation testing} -- a form of randomized experiments, where individuals with the same characteristics except for the protected ground are sent to apply for a job or for a loan.
Consider the contingency table in Figure~\ref{symbolic} for the set of $k$-nearest neighborhood tuples $\mathit{kset}_{\mathcal R}({\bf r}, k)$ of ${\bf r}$. Here, individuals are modelled by the vector of their features, and a similarity function between tuples is assumed to be defined.
The risk difference $RD$ defined in Figure~\ref{symbolic} is a measure of the degree of bias towards negative decisions for members of ${\bf r}$ to the protected group. If such a degree is above a threshold $\alpha \geq 0$, i.e.,~$RD > \alpha$, then the individual ${\bf r}$ was discriminated. The threshold $\alpha$ is a parameter of the analysis A dual statement for an individual ${\bf r}$ of the unprotected group with a positive decision,i.e.~$\mathit{dec}({\bf r}) = \oplus$, is the question: was ${\bf r}$ favored?

While we concentrate on $RD$, our approach readily applies to other measures defined in terms of contingency tables such as risk ratio, odds ratio, etc., and to their tests of
statistical significance (see surveys~\cite{RR2014,DBLP:journals/datamine/Zliobaite17}).
A major drawback of directly using a discrimination measure over the contingency table as in Figure~\ref{symbolic} is that it does not account for differences between the individuals in the neighborhood. Although we consider in the neighborhood tuples that are close to each other, differences in the attribute values of tuples in the protected and those in the unprotected group may still support different decisions for such groups. For example, in an employee salary dataset several factors may contribute to the high/low salary of an employee (education, work hours, age, etc.). While it is desired that males and females with similar credentials should have similar salaries, observed difference in salaries of males and females in the neighborhood of an employee could signify potential discrimination or favoritism. The calculated risk difference of the employee, however, would be biased if is found, for instance, that females in the dataset work fewer hours than males and employees working for fewer hours obtain lower salaries. Thus, the factor work hours confounds the estimated risk difference. We aim, instead, at estimating the \emph{causal} effect of belonging to the protected group on the outcome by handling the effect of confounding factors.

\textbf{Propensity Score Analysis.} Our problem commonly arises in the statistical analysis of observational data where assignment of individuals to treatment and control groups cannot be assumed to be at random~\cite{shadish2002}. In our context, we consider the protected group as the treatment group, and the unprotected group as the control group.
A method for causal analysis over observational data is provided by propensity scores. Propensity score analysis is a principled way to handle multiple confounding variables and to `filter out' the explainable effect of these variables.

\begin{definition}
Assume that the dataset ${\mathcal R}$ is a sample over a distribution of covariates ${\bf x}$. 
The propensity score $e(\cdot)$ is the conditional probability of belonging to the protected group given the covariates: \[ e({\bf x}) = Pr(\mathit{protected} \ |\ {\bf x} ) \]
\end{definition}

The estimation of propensity scores requires two key decisions: the model or functional form of $e(\cdot)$ and the (confounding or explanatory) variables to include in ${\bf x}$. For binary valued groups, as in our case, the logistic regression model is commonly adopted ~\cite{caliendo2008some}. 
The decision regarding which covariates to include in the logistic regression model is less standardized. It is recommended that the selected variables should influence simultaneously the group membership and the decision of individuals. Furthermore, it is advised against selecting too many covariates such that estimates of propensity scores for most individuals becomes exactly one or zero which violates the condition of overlap between protected and unprotected group individuals w.r.t. propensity score necessary for effective propensity score analysis~\cite{bryson2002use}. Following these guidelines, we employ the following estimation procedure:

\textit{1)} The propensity score is given by a logistic regression model learned over the dataset $\mathcal{R}$, i.e.,
\[
e(\mathbf{x}) = \frac{1} {1 + \exp(-\mathbf{\beta}^{\rm T}
\mathbf{x})}
\]
\noindent Here, $\mathbf{x}$ is the vector of selected covariates and $\mathbf{\beta}$ is the corresponding weight vector.

\textit{2)} The variables that are correlated with both group membership and decision are candidates for inclusion in the propensity score model. The correlation measure and its threshold is decided by a domain expert. For example, variables producing information gain $\geq 0.05$ w.r.t. both group and decision variables could be candidates for selection.
 
\textit{3)} Remove from the selected candidates, variables that are proxies for the group membership, i.e., variables that are almost duplicates of the group variable, as they can cause the propensity scores to be either zero or one. For example, variables producing an entropy $\geq 0.95$ for predicting the group variable are removed.

Propensity scores can answer the following counter-factual question: ``what kind of outcomes would we have observed had the decision involving unprotected individuals involved protected ones instead?" Let us introduce now weights $w({\bf x})$ such that the distribution of the protected and unprotected group becomes identical, i.e.,~such that:
\[ Pr( {\bf x} \ |\ \mathit{protected} ) = w({\bf x}) \cdot Pr( {\bf x} \ |\ \mathit{unprotected}
)\]
By simple algebra and the Bayesian rule, we obtain:

\begin{equation}\label{def:wx}
w({\bf x}) = K \cdot \frac{e({\bf x})}{1 - e({\bf x})}
\end{equation}
where $K = Pr(\mathit{unprotected})/Pr(\mathit{protected})$ is a constant term that will not be relevant in subsequent analysis. $w({\bf x})$ is the weight that a tuple ${\bf x}$ of the unprotected group should count for in case it would belong to the protected group. This approach is known as \emph{propensity score weighting}~\cite{Rosenbaum83}. Rosenbaum and Rubin showed that, conditional on the propensity score, all observed covariates are independent of group assignment and, in large samples, they will not confound estimated treatment effects, i.e.,~in our context the decisions on the protected group can be compared with the decisions on the unprotected group once the latter is re-weighted using propensity scores.

\textbf{Propensity Score Weighting.} Let us now apply propensity score weighting in the context of discrimination measures over a contingency table for an individual ${\bf r}$ of the protected group. See Figure~\ref{symbolic}. Instead of comparing the average negative decision $p_1$ of the protected group with the average negative decision $p_2$ of the unprotected group, we compare $p_1$ with the average negative decision $p_2^c$ that would be obtained for individuals of the unprotected group if they were instead in the protected group. We define the weighted average negative decision of the unprotected group $p_2^c$ as:
\[ p_2^c = \frac{\sum_{{\bf s} \in {\mathcal S}, \mathit{dec}({\bf s}) = \ominus} w({\bf s}) }{\sum_{{\bf s} \in {\mathcal S}} w({\bf s})}\] where ${\mathcal S} = \mathit{kset}_{\mathcal R}({\bf r}, k, m) \cap {\mathcal
R}(\mathit{unprotected})$ is the set of tuples in the neighborhood that belong to the unprotected group. Notice that the constant term $K$ in (\ref{def:wx}) is ruled out in the calculation of the ratio $p_2^c$.

\begin{definition}\label{def:rdc}
The causal risk difference is $RD^c = p_1 - p_2^c$.
\end{definition}

The causal risk difference compares the proportion of negative decisions for the neighboorhood tuples in protected group with the proportion of negative decisions for the neighboorhood tuples of the unprotected group re-weighted to the distribution of the protected group. 
$RD^c$ close to zero means that differences in the outcomes of the decision between the protected and unprotected group are motivated by confounding factors in the selection of the two groups which are covered by the covariates of the propensity scores. If $RD^c$ is not close to zero, then there is a (causal) bias in decision value due to group membership, or there are covariates that have been not accounted for in the analysis (\emph{omitted variables}). In the former case, $RD^c > 0$ denotes causal discrimination, and $RD^c < 0$  denotes causal favoritism (e.g.,~due to positive/affirmative actions).

\begin{example}
Let us contrast on a fictitious example risk difference and causal risk difference with the variation of propensity score. Fix a tuple of the protected group \textit{with a negative decision}, and fix $k = 15$. Suppose the set of the $k$ nearest neighbors of this individual contain 7 females, with 4 having a negative decision, and 8 males, with 3 having a negative decision. The (non-causal) risk difference of this individual is $RD = p_1 - p_2 = 4/7 - 3/8 = 0.196$. Suppose that males have a propensity score of $0.5$. Their weights will be $0.5/(1-0.5) = 1$ and then $p^c_2 = p_2$, hence $RD^c = RD$. Intuitively, if there is no difference in the distribution of males and females, causal risk difference boils down to risk difference. Assume now that the 5 males with positive decision have a greater propensity score, i.e., their characteristics are closer to the female group distribution than to the one of the male group. Then their weights will become greater than $1$ and $p^c_2 < p_2$, hence $RD^c > RD$. For instance, when the propensity score of the positive males is $0.8$, the weight for the positive males is equal to $4$, and the causal risk difference of the individual becomes $RD^c = 4/7 - 3/23 = 0.441 > RD$. That is, the individual faces higher discrimination than the one quantified by risk difference, because similar males are given positive decisions even though they have `femalish' characteristics. The opposite case is also possible, namely $RD^c < RD$, i.e.,~non-causal risk difference overestimates discrimination. For example, if the 5 males with positive decision have propensity score of $0$, i.e., they are definitely not following the distribution of females, their weight is also $0$, and then $RD^c = 4/7 - 3/3 = -0.429 < RD$.
\end{example}

\section{Causal Discrimination Discovery}
\label{causaldiscriminationdiscovery}

In the previous section, we assigned to every tuple ${\bf r}$ of the protected group with negative decision a causal measure $RD^c$ of discrimination (for $RD^c > 0$) observable when comparing its decision to that of its neighbors. For tuples of the unprotected group, a similar measure can be defined as $RD^c = p_2^c - p_1$, so that again $RD^c > 0$ denotes discrimination. Dually, $RD^c < 0$ denotes favoritism for individuals of the protected and unprotected groups experiencing a positive decision.
This approach is useful for quantifying bias in discriminating/favoring an individual (\textit{individual discrimination}). However, in many cases, a global description of who was discriminated/favored is also required. In order to infer such descriptions, we proceed by extracting a regression model from a modified version of the dataset ${\mathcal R}$ at hand. Consider the problem of characterizing discrimination in the protected group. We restrict to tuples of the protected group with negative decision. We augment the dataset with a new attribute, the class attribute, setting its value to the $RD^c$ for each tuple in it. Over such a labeled dataset, a regression tree~\cite{Brei84} is extracted, which can be used for descriptive purposes. A path in the regression tree ending in a prediction of (large) positive values describes a context of discrimination for the protected group. Similarly, patterns of favoritism for the unprotected group with positive decision can be studied by extracting regression trees and looking at paths predicting large negative values of the class. In summary, we adopt regression to lift from causal individual discrimination analysis to causal global discrimination analysis.

\begin{figure}[t]
\centering
\small
\includegraphics[width=0.45\textwidth]{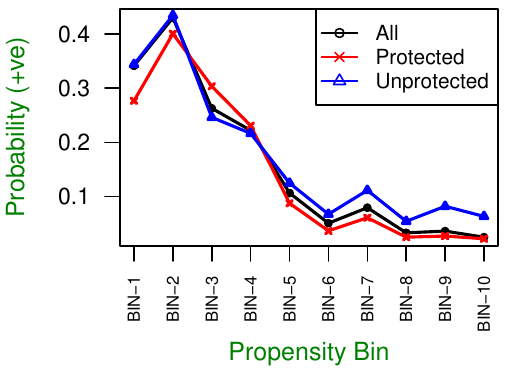} \hspace{5ex}
\includegraphics[width=0.45\textwidth]{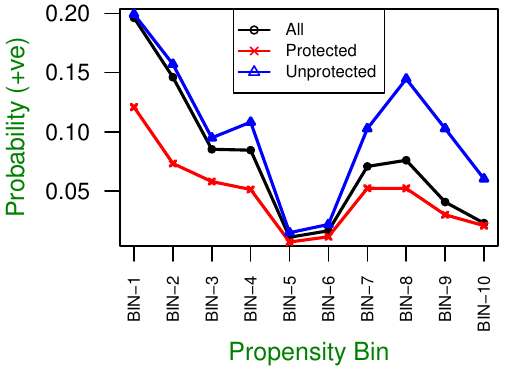}\\
(a) Adult: $\oplus$ class prob. \hspace{20ex}
(b) Census: $\oplus$ class prob.
\\
\caption{Variation of positive decision probability with propensity score.}
\label{fig:positive-probability}
\end{figure}
\section{Experiments}
\label{sc:exp}
We perform experiments\footnote{Code at: \href{https://github.com/anam-zahid/Causal-inference-for-social-discrimination-reasoning}{https://github.com/anam-zahid/Causal-inference-for-social-discrimination-reasoning}} on two commonly-used datasets in discrimination analysis research: Adult, and Census Income. Both datasets are publicly available in the UCI Machine Learning Repository\footnote{\href{http://archive.ics.uci.edu/ml}{http://archive.ics.uci.edu/ml}}. In both datasets, we use \code{income} as the decision attribute with \code{income $>$ 50K} as the positive decision and \code{sex} as the sensitive attribute considering females as the protected group.

Propensity scores are estimated by a logistic regression model learned over the dataset for predicting membership to the protected group. The attributes used for propensity score estimation are those correlated with both decision and group attributes. Information gain is used as the correlation measure and the selection threshold is set to 0.05.

The neighborhood of a tuple (namely, $\mathit{kset}_{\mathcal R}({\bf r}, k)$) is defined by setting $k = 15$. The Euclidean distance with attributes normalized (min-max normalization) in the interval [0, 1] is used as the distance function.

\textbf{Variation of Positive Decisions with Propensity Scores.} To study the variation of positive decisions with propensity scores, we discretized the datasets into 10 equal interval bins on the basis of propensity scores, e.g., $\emph{bin-1}$ instances have propensity (of protected group membership) less than 10\%, $\emph{bin-2}$ with 10-20\% and it ends with $\emph{bin-10}$ having propensity score between 90-100\%.  Individuals of bins 1-5 have characteristics that are more similar to those in the unprotected group (males) due to their low propensity of belonging to the protected group (females).

Figure~\ref{fig:positive-probability} shows the difference in positive class probabilities of tuples based on their propensity score bins and group membership.

It turns out that:
instances with male-like characteristics, i.e., bins 1-5, have considerably higher $\oplus$ decision probabilities than female-like bins; and, that $\oplus$ decision probability difference between protected and unprotected groups is also in favor of unprotected group instances in most of the bins. Differences are more marked for the Census dataset compared to the Adult dataset.

\begin{figure}[t]
\centering
\small
\hspace{-2ex}
\includegraphics[width=0.45\textwidth]{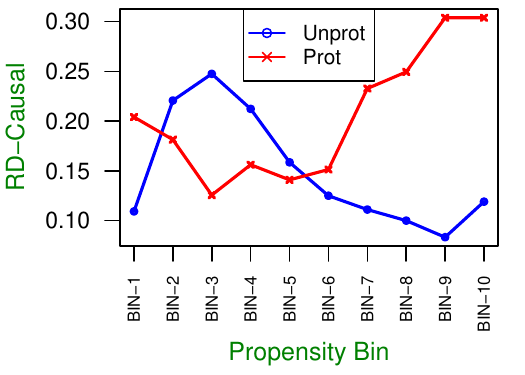}  \hspace{5ex}
\includegraphics[width=0.45\textwidth]{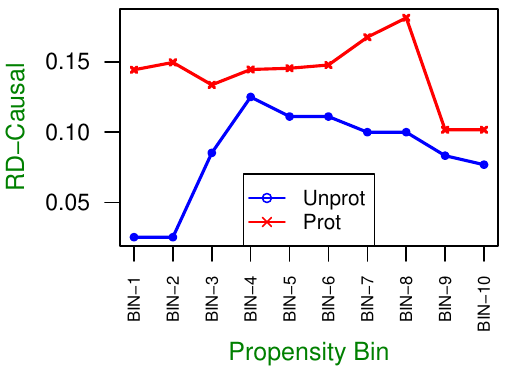} \\
(a) Adult: discrimination. \hspace{20ex}
(b) Census: discrimination.\\
\hspace{-4ex}

\includegraphics[width=0.45\textwidth]{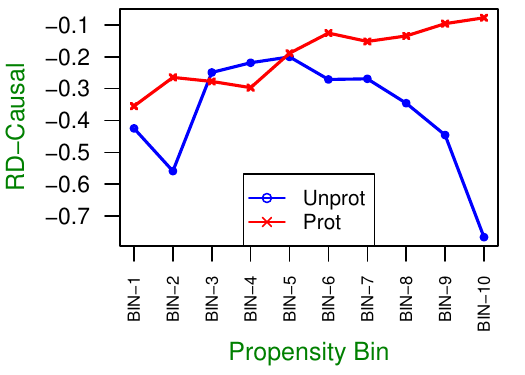} \hspace{5ex}
\includegraphics[width=0.45\textwidth]{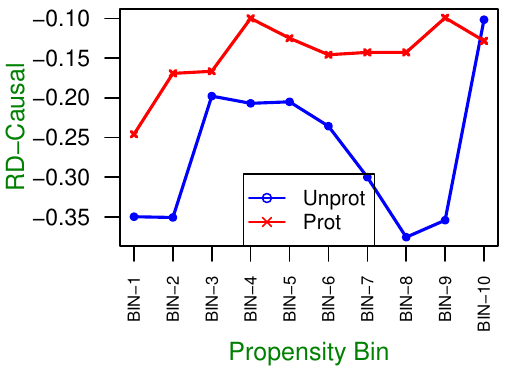}\\
(c) Adult: favoritism. \hspace{22ex}
(d) Census: favoritism.\\
\caption{Group wise discrimination/favoritism trends.}
\label{fig:group-comp}
\end{figure}

\textbf{Group Wise Discrimination/Favoritism Trends.} Plots in Figure~\ref{fig:group-comp} show group wise discrimination/favoritism trends measured through  causal risk difference at the variation of propensity score. Propensity bins are on the x-axis while average $RD^c$'s are on the y-axis. Top row plots contain instances with positive $RD^c$ (discriminated) while bottom row plots contain instances with negative $RD^c$ (favored).

General assumption is that protected group instances are more discriminated than unprotected group instances. Figure~\ref{fig:group-comp} (a) support this assumption by showing that protected group instances face more discrimination in most of the propensity bins except with a few exceptions in male-like bins 2-5. Such a group-grounded bias becomes more evident in the Census Income dataset, as shown in Figure~\ref{fig:group-comp} (b). Fig \ref{fig:group-comp} (c \& d) show a significant difference in the favoritism scores between both groups for both datasets. Unprotected group individuals (blue line) are highly favored in most of the propensity bins despite of having characteristics quite similar to protected instances. Fig \ref{fig:group-comp} carries a lot of importance and shows very obvious discrimination/favoritism trends between both sensitive groups despite of similar propensity scores and characteristics. 


%
\begin{figure}[t]
\centering
\small
\hspace{-2ex}
\includegraphics[width=0.45\textwidth]{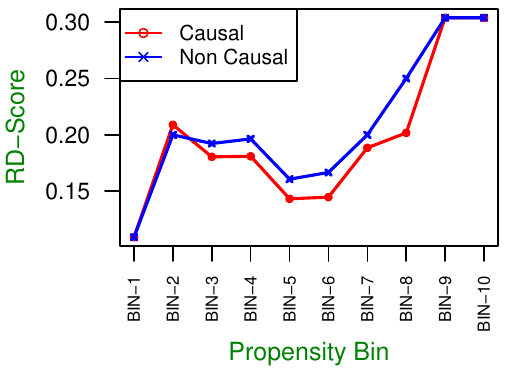} \hspace{5ex}
\includegraphics[width=0.45\textwidth]{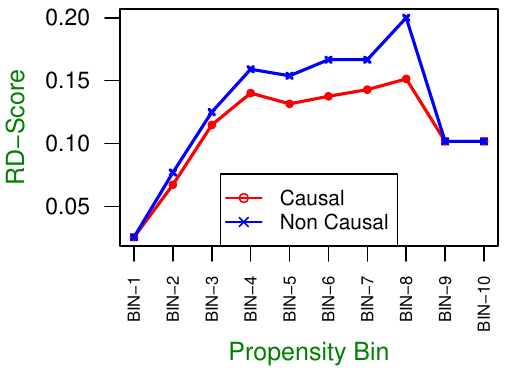}\\
(a) Adult: discrimination.  \hspace{20ex}
(b) Census: discrimination. \\
\hspace{-4ex}

\includegraphics[width=0.45\textwidth]{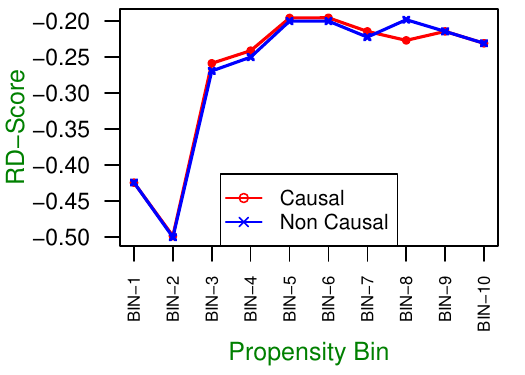} \hspace{5ex}
\includegraphics[width=0.45\textwidth]{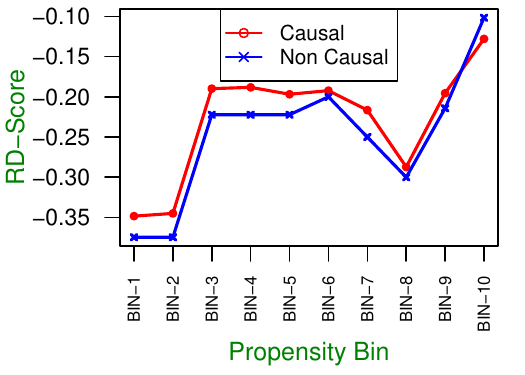}\\
(c) Adult: favoritism.  \hspace{20ex}
(d) Census: favoritism. \\
\caption{$RD$ and $RD^c$ comparison.}
\label{fig:causal-comp}
\end{figure}

\textbf{$RD$ and $RD^c$ Comparison.} Figure~\ref{fig:causal-comp} shows a comparison between $RD^c$ and its non causal variant $RD$. According to our discussion earlier in Section \ref{causalindividualdiscrimination}, $RD$ overestimates both discrimination and favoritism scores (recall that higher absolute values denote higher discrimination/favoritism). Margin between $RD$ and $RD^c$ lines show causation magnitude. 
%
Top row shows discriminated instances with positive $RD$ and $RD^c$  and bottom row represents favored instances with negative scores. 
Figure~\ref{fig:causal-comp} (a) shows higher discrimination scores by $RD$. Similar trends are seen in Figure~\ref{fig:causal-comp} (b) and (d) while $RD$ and $RD^c$ are relatively closer in Figure~\ref{fig:causal-comp} (c). We can conclude that $RD$ is overestimating discrimination/favoritism scores in almost all the bins of both datasets but the margin is higher for the Census Income dataset. 

\begin{figure*}
\centering\small
\begin{tabular}{cc}
\hspace{-5ex}\includegraphics[width=0.54\textwidth]{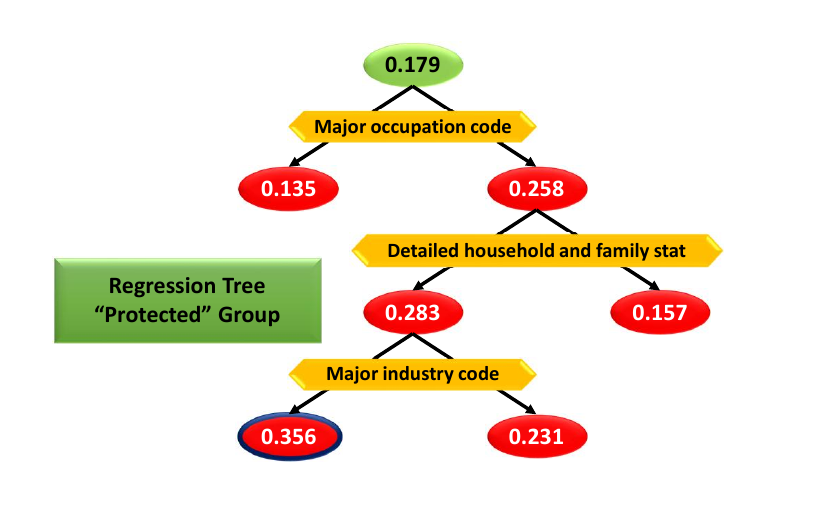} &
\hspace{-5ex}\includegraphics[width=0.54\textwidth]{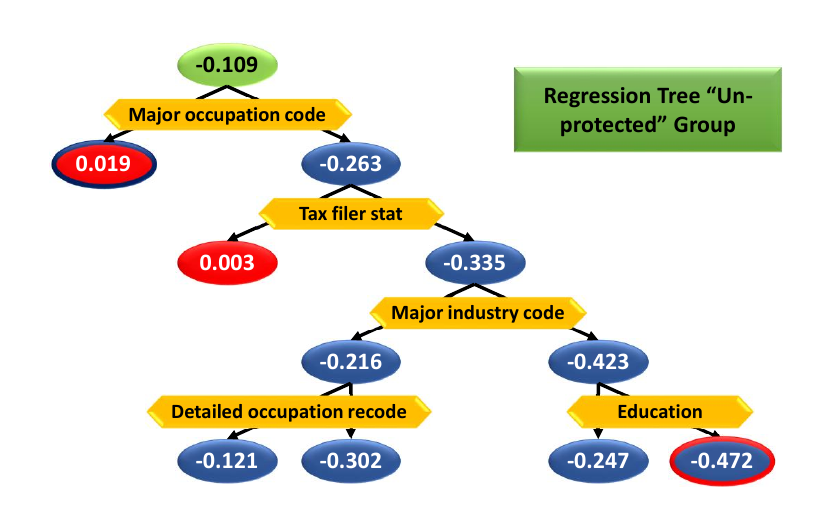} \\
(a) Discrimination against the protected group. &
(b) Favoritism for the unprotected group.\\
\end{tabular}\\
\caption{Census Income: regression trees.}
\label{fig:pattern-identification}
\end{figure*}

\textbf{$RD^c$ and Rule Generation.}
We use regression rules to explore interesting patterns in the datasets that lead to high values of discrimination/favoritism scores. Rules are created from protected and unprotected groups separately and only include instances with non-zero $RD^c$. In our analysis, every node (whether leaf or not) acts as a rule and its label (the average risk difference of tuples reaching the node) acts as bias magnitude. Figure~\ref{fig:pattern-identification} (a \& b) 
belong to protected and unprotected groups of the Census Income dataset. 
The green node is the root, red nodes show discriminatory rules while blue nodes show favoritism rules. Nodes with maximum discrimination/favoritism magnitude are highlighted with contrasting borders. 

%

In Figure~\ref{fig:pattern-identification} (a), after the split on \code{major occupation code}, we get two subsets with 13.5\% and 25.8\% average causal risk difference respectively. Similarly we discover multiple other subsets that have different magnitudes of discrimination/favoritism. The most discriminated group suffers from a 35.6\% causal risk difference. Similar trends are seen in Figure~\ref{fig:pattern-identification} (b) where most of nodes regard favoring sub-groups of unprotected individuals. E..g, in the rightmost node, males with comparable characteristics as females benefit from high income with a causal risk difference of -47.2\%. This is a key observation of our causal discrimination discovery technique that highlights under which conditions (the path from root to the node) discrimination and favoritism do arise.

\section{Conclusions}
\label{conclusions}

Based on propensity score analysis, our approach for discrimination discovery and understanding constructs a modified distribution of observed covariates that eliminates the effect of observed confounding factors. The degree of discrimination or favoritism is quantified by the risk difference computed over the modified distribution. 
This is the starting point for measuring and understanding discrimination bias in training data (preventing discrimination learning) or in test data (assessing discrimination in predictive models). While graphical/structural causation models are able to model finer-grained discrimination analyses, our approach does not require human intervention, and scales linearly with the number of features. This makes it feasible to apply the approach in practice. Moreover, our approach can be easily implemented on top of basic data analysis tools (logistic regression, nearest-neighbor set, regression trees).
Future work is open towards several directions. A first one is to adapt training data sanitization~\cite{DBLP:conf/kdd/ThanhRT11} and fair machine learning models~\cite{Berk2017} to account for causal discrimination measures. A second possible direction is to extend our approach and experimentation to non-structured and high-dimensional data types which may also hide human biases, such as text~\cite{DBLP:conf/nips/BolukbasiCZSK16}, which originate in the web~\cite{DBLP:journals/cacm/Baeza-Yates18} and from social media~\cite{DBLP:journals/ir/KulshresthaEMZG19}.

%
%
%
\bibliographystyle{splncs04}
\bibliography{causal.bib}
\end{document}